\documentclass{aa}

\usepackage{natbib}

\begin{document}

\title{Discovery of magnetic fields in hot subdwarfs\thanks{Based on
    observations collected at the European Southern Observatory, Paranal,
    Chile, under programme ID 072.D-0290(A).}}

\author{S.~J. O'Toole
  \inst{1}
  \and S. Jordan
  \inst{2}
  \and S. Friedrich
  \inst{3}
  \and U. Heber
  \inst{1}
}

\institute{Dr.\ Remeis-Sternwarte, Astronomisches Institut der
  Universit\"at Erlangen-N\"urnberg, Sternwartstr.\ 7, Bamberg D-96049,
  Germany
  \and Astronomisches Rechen-Institut, M\"onchhofstr.\ 12-14, 69221
  Heidelberg, Germany
  \and Max-Planck-Institut f\"ur extraterrestrische Physik, Giessenbachstr.,
  85748 Garching, Germany
}

\offprints{Simon O'Toole,
\email{otoole@sternwarte.uni-erlangen.de}}

\date{Received / Accepted}

\abstract{We present polarisation measurements of sdB and sdO
  stars using \emph{FORS1} on the \emph{VLT}. The observations were made as
  part of a project to determine whether magnetic fields in two
  super-metal-rich stars can explain their extreme abundance
  peculiarities. Field strengths of up to $\sim$1.5\,kG range have been
  measured at varying levels of significance in each of our six targets,
  however no clear evidence was found between apparently normal subdwarfs and
  the metal-rich objects. The origin of the magnetic fields is unknown. We
  also discuss the implications of our measurements for magnetic flux
  conservation in late stages of stellar evolution.
}

\titlerunning{Magnetic fields in hot subdwarfs}
\authorrunning{S.~J. O'Toole et al.}

\maketitle

\section{Introduction}
\label{sec:intro}
The measurement of magnetic fields in evolved stars at the kilogauss
level has become possible only recently, when spectropolarimeters attached to
telescopes with diameters larger than 8\,m have become available.
The implementation of circularly-polarised spectropolarimetry with
\emph{FORS1} spectrograph of the Very Large Telescope (\emph{VLT}) has opened
the door to study magnetic fields in a large range
of stars. The technique was first described by \citep{BSW02}, who used the
hydrogen Balmer lines to measure a mean longitudinal field of $\sim$2\,kG in a
chemically peculiar A star.
Such low magnetic fields have now been detected in white dwarfs and central
stars of planetary nebulae \citep{AJN04,JWO05}.

Hot subdwarfs are subluminous objects that dominate the population of faint
blue stars in our own galaxy. The subdwarf B stars (sdBs) have hydrogen-rich
atmospheres with effective temperatures below about 40\,000\,K
\citep[e.g.][]{Heber86}. They are also typically helium-poor. Subdwarf O (sdO)
stars on the other hand cover a much larger range of atmospheric compositions
with a large spread of hydrogen and helium abundances. Their effective
temperatures are in the range between 40000 and 90000\,K. In the $\log
g-T_{\rm eff}$ diagram they are found close to (``born-again'') post-AGB or
post-extreme horizontal branch tracks \citep[e.g. ][]{HBH89, SHL05}. Both
groups are believed to be the progeny of 1-2\%\ of the white dwarfs.
Recently, two subdwarf B stars (UVO\,0512--08 and PG\,0909+276) were
discovered to have unusually high abundances of iron-group elements, e.g.\
these stars have up to $10^5$ times more vanadium and nickel than the Sun
\citep{EHN01}. While no direct evidence was found in their spectra, one
obvious possibility is that these anomalously high abundances are somehow
related to large magnetic fields, as seen in the chemically peculiar A
stars. Inspired by this discovery, we started a project to measure magnetic
fields in hot subdwarfs.

\begin{table*}
\caption{List of targets observed; super-metal-rich objects are marked with an
  asterisk.}
\label{tab:targets}
\begin{center}
\begin{tabular}{cccccccccccc}
Target & $\alpha$ & $\delta$ & $V$ & UT & $t_{\mathrm{exp}}$ & $n$ &
 $T_{\mathrm{eff}}$ & $\log g$ & log(He/H) & sp.\ type & Ref. \\
 & (J2000) & (J2000) & (mag) & start & (s) & (K) & & & \\
\hline
UVO\,0512--08\rlap{$^*$} & 05 14 44.0 & $-$08 48 04 & 11.3 & 00:10:56 & 440 & 3 & 38\,800 & 5.5 &
 $-$0.8 & sdB & 1 \\
CPD\,$-64^\circ 481$ & 05 47 59.3 & $-$64 23 03 & 11.3 & 01:14:00 & 440 & 3 & 27\,500 &
 5.6 & $-$2.9 & sdB & 2 \\
PG\,0909+276\rlap{$^*$} & 09 12 51.7 & +27 20 31 & 12.1 & 02:16:27 &1200 & 3 & 36\,900 & 5.9 & $-$0.8 & sdB & 1 \\
CD\,$-31^\circ 4800$ & 07 36 30.2 & $-$32 12 44 & 10.6 & 04:54:30 & 240 & 3 & 44\,000 & 5.9 & $\sim$1.5
 & sdO & 4 \\
HD\,76431 & 08 56 11.2 & +01 40 38  & 9.2 & 05:39:56 &110 & 9 & 31\,000 & 4.5 & $-$1.5 & sdB & 3 \\
LSE\,153 & 13 53 08.2 & $-$46 43 42  & 11.4 & 06:44:52 & 550 & 9 & 70\,000 & 4.75 & $\sim$2 &
 sdO & 5 \\
\hline
\end{tabular}
\end{center}
References: (1) \citet{EHN01}; (2) This work; (3) \citet{RHE01}; (4)
\citet{BH95}; (5) \citet{HBH89}.
\end{table*}

There is only one published study of a search for magnetic fields in sdB and
sdO stars, which was carried out using the Special Astronomical Observatory
6\,m telescope \citep{Elkin1996}.
The author inferred a longitudinal  magnetic field of $-1690\pm 60$\,G (based
on three measurements of the polarisation in the He\,\textsc{i} 5875\,\AA\
line of the sdO BD\,+75$^\circ$325 ($V=9.21$) and a variable field in the
sdOB Feige 66 ($V=10.64$) with measurements of  $-1300\pm 320$\,G
(He\,\textsc{i} 5875), $-1300\pm 580$ (He\,\textsc{i} 5875), and $+1750\pm
230$\,G (H$\alpha$, He\,\textsc{i} 5875), and $+460$\,G (H$\alpha$), each
value based on one or two measurements. The uncertainties  were (probably)
calculated from the deviations of successive observations without taking into
account individual errors in measurements or sytematic errors.

In this paper we present spectropolarimetric observations of sdBs and
sdOs based on polarisation measurements using the Balmer lines and
He\,\textsc{i} and \textsc{ii} lines.


\section{Observations and Reduction}
\label{sec:obsred}

Circular polarisation spectra of six hot subdwarfs were obtained using
the VLT/UT1 (Antu) with the FORS1 spectrograph on February 23, 2004 (UT) in
visitor mode. FORS1 is a multi-mode focal reducer imager and grism
spectrograph equipped with a Wollaston prism and rotatable retarder plate
mosaics in the parallel beam allowing linear and circular polarimetry and
spectropolarimetry. We used the 600B grism, covering the spectral range
3400--5900\,\AA, with a 0.5'' slit, leading to a spectral resolution of
2.8\,\AA. Details about our targets can be found in Table
\ref{tab:targets}. Two out of six of them are the super-metal-rich sdB stars
found by \citet{EHN01}, one is an apparently normal sdB, one is a post-EHB
sdB, while the last two stars are sdOs. The two metal-rich objects were
observed in the first half of the night, and subsequent bad weather meant that
more ``normal'' abundance targets could not be observed.

Bias frames, flat-fields and He+HgCd arc spectra were taken at the end of the
night. The data were reduced using standard \textsc{iraf} routines for bias
subtraction, cosmic ray removal and flat fielding.
After the flat-field correction, the stellar spectra were
extracted from each frame by summing up all CCD rows for the $e$ and $o$
beams. Background sky light was averaged over 10 rows on either side of the
object spectrum (giving a total of 20 rows) and subtracted.

\begin{figure*}
\vspace{18cm}
\begin{center}
    \includegraphics{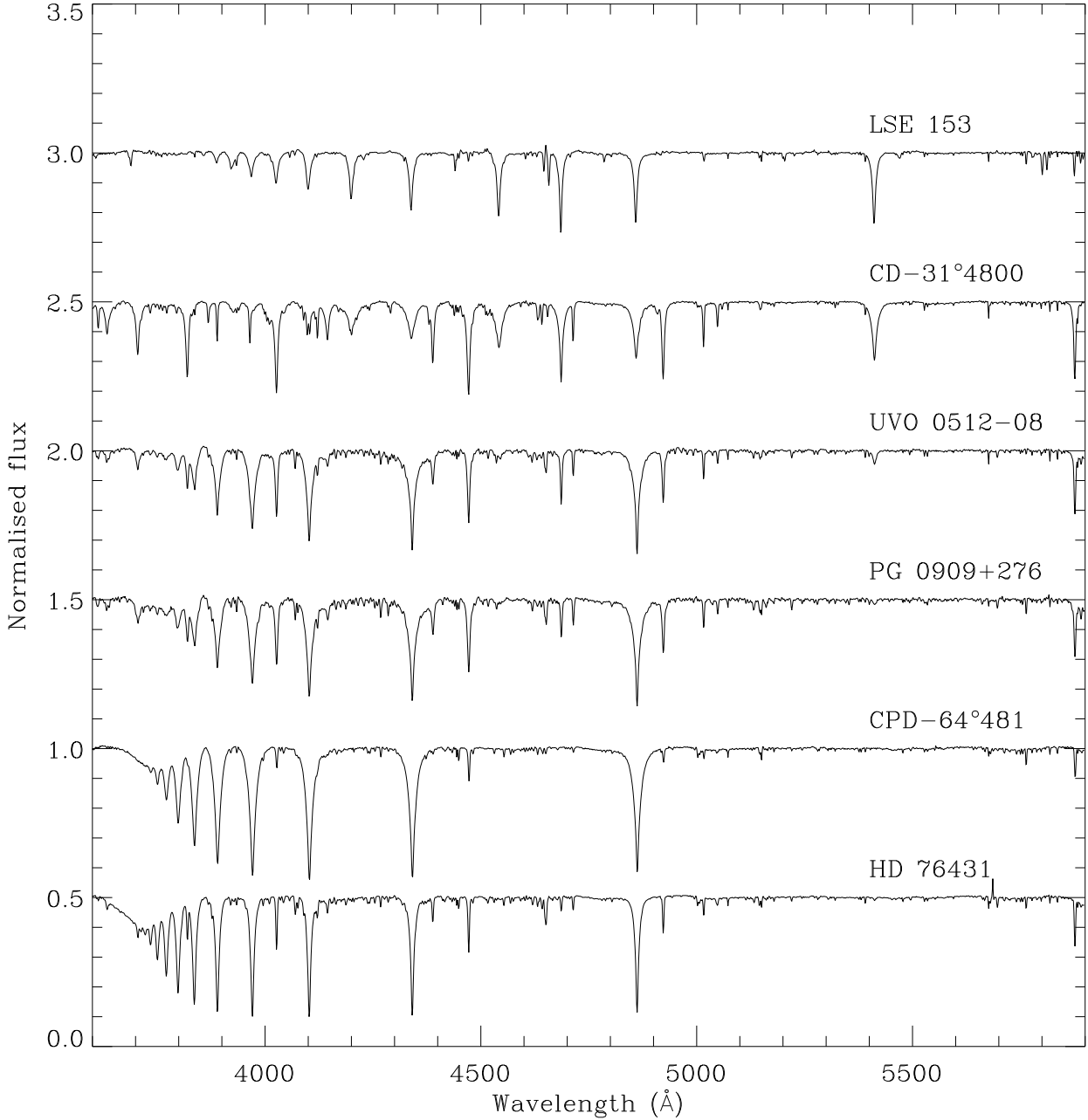}
\end{center}
\caption{Normalised spectra of our six targets. The spectra are offset by 0.5
  for clarity.}
\label{fig:normspec}
\end{figure*}

As noted by \citet{JWO04}, it is very important to make sure that the spectra
are correctly wavelength calibrated; an incorrect calibration might lead to
a spurious polarisation signal.
Calibration was done independently for the spectra of each beam ($e$ and $o$)
and each position of the retarder plate ($\pm 45^\circ$). The wavelength
accuracy of each spectrum is about 3\,km\,s$^{-1}$ or 0.05\,\AA\ at H$\beta$,
which is much lower than the spectral resolution. Stokes \emph{I} spectra
(i.e. unpolarised, the sum of the $e$ and $o$ beams) are shown in Figure
\ref{fig:normspec} with the continuum normalised to unity and each spectrum
offset by 0.5 (not scaled) for clarity. Note that the top two spectra are the
helium-rich sdO stars, the two middle spectra are the super-metal-rich sdBs
and those at the bottom are the ``normal'' abundance sdBs.

To derive the level of circular polarisation from such a sequence, we added
the exposures taken with the same quarter-wave plate angle. Stokes $V/I$ can
then be found using
\begin{equation}
\frac{V}{I}=\frac{R-1}{R+1},\; \mathrm{with}\: R^2=\left (\frac{f_o}{f_e}
\right )_{\alpha=+45} \times \left (\frac{f_e}{f_o} \right )_{\alpha=-45}
\label{eq:VonI}
\end{equation}
where $V$ is the Stokes parameter describing the net circular polarisation,
$I$ is the unpolarised intensity, $\alpha$ indicates the nominal value of the
position angle of the retarder-wave plate, and $f_o$ and $f_e$ are the fluxes
on the detector from the ordinary and extraordinary beams of the Wollaston
prism, respectively.

The mean longitudinal field can be derived using the weak field approximation,
which is valid in the presence of instrumental broadening, but not in the case
of rotational broadening. The limiting value of $v\sin\,i$ is dependent
  on the intrinsic line width; the Balmer lines are not expected to be
  affected because Stark broadening is dominant. For example, if the line
  width is $\sim$15\,\AA\ (typical for the Balmer lines in hot subdwarfs),
  $v\sin i$ would have to be greater than 1000\,km\,s$^{-1}$, while a narrow
  line with width 3\,\AA\ would be useless if $v\sin i>200$\,km\,s$^{-1}$.
None of our targets are known to be rapid rotators in any case. The
field strength is compared with the Stokes $V/I$ spectrum using the following
equation \citep{Landstreet82}:
\begin{equation}
\frac{V}{I}=-g_\mathrm{eff}\frac{e}{4\pi
  m_ec^2}\lambda^2\frac{1}{I_\lambda}\frac{dI}{d\lambda}\langle
  B_z\rangle
\label{eq:weak}
\end{equation}
where $g_{\mathrm{eff}}$ is the effective Land\'e factor, $\lambda$ is
wavelength in \AA, $\langle B_z\rangle$ is the mean longitudinal
field in Gauss, and $e/(4\pi m_ec^2)\simeq 4.67
\times10^{-13}$\,G$^{-1}$\,\AA$^{-1}$. Note that we are only measuring the
longitudinal field average over the stellar surface; the maximum field
strength may be larger.

The longitudinal component of the magnetic field for each measurement was
determined by comparing the observed circular polarisation for an interval of
$\pm 20$\,\AA\ around the strongest absorption lines with the prediction of
Equation\,\ref{eq:weak}. As in \cite{AJN04} and \citet{JWO04}, we determined
$\langle B_z\rangle$ using a $\chi^2$-minimisation procedure. Unlike the case
of central stars of planetary nebulae, systematic errors due to the blending
of Balmer lines and the Pickering series He\,\textsc{ii} lines should be
small. This is because in the stars where the hydrogen lines are strong, the
helium lines are not strong, and vice versa. In the case of strong Balmer
lines, some of the helium lines are very weak. The strongest are
He\,\textsc{i} 5875, He\,\textsc{i} 4471, He\,\textsc{i} 4921 and
He\,\textsc{ii} 4686. We used the criterion that the line must be stronger
than 10\% below the continuum to be measurable. Because of this, only in the
two He-rich sdO stars could other helium lines be used.
Other systematic errors besides the blending effect are difficult to estimate
in strength, but we believe that they are lower than 200\,G.

\begin{table*}
  \caption[]{Effective Land\'e factors for the observed He\,\textsc{i} and
  He\,\textsc{ii} lines, along with the next three strongest lines that may be
  observed with a different \emph{FORS1} grism.}
  \label{tab:lande}
\begin{center}
  \begin{tabular}{l||c||c|c|c|c||c|c|c|c}
    \hline
\multicolumn{10}{c}{He\,\textsc{i}}\\
\hline
  line&$g_{\mathrm{eff}}$&\multicolumn{4}{|c||}{lower
  level}&\multicolumn{4}{|c}{upper level}\\
&&$J$&$L$&$S$&$g$&$J$&$L$&$S$&$g$\\
\hline
4026.29\,\AA&1.17&2&1&1&3/2&3&2&1&4/3\\
4471.60\,\AA&1.17&2&1&1&3/2&3&2&1&4/3\\
4713.28\,\AA&1.25&2&1&1&3/2&1&0&1&2\\
4921.94\,\AA&1.00&1&1&0&1&2&2&0&1\\
5015.69\,\AA&1.00&0&0&0&--&1&1&0&1\\
5047.74\,\AA&1.00&1&1&0&1&0&0&0&--\\
5875.83\,\AA&1.17&2&1&1&3/2&3&2&1&4/3\\
6678.16\,\AA&1.00&1&1&0&1&2&2&0&1\\
7065.53\,\AA&1.25&2&1&1&3/2&1&0&1&2\\
7281.37\,\AA&1.00&1&1&0&1&0&0&0&--\\
\hline
\hline
\multicolumn{10}{c}{He\,\textsc{ii}}\\
\hline
  line&$g_{\mathrm{eff}}$&\multicolumn{4}{|c||}{lower
  level}&\multicolumn{4}{|c}{upper level}\\
&&$\bar{J}$&$L$&$S$&$g$&$\bar{J}$&$L$&$S$&$g$\\
\hline
4100.05\,\AA&1.06&3.5&3&0.5&1.14&4.5&4&0.5&1.1\\
4199.84\,\AA&1.06&3.5&3&0.5&1.14&4.5&4&0.5&1.1\\
4338.68\,\AA&1.06&3.5&3&0.5&1.14&4.5&4&0.5&1.1\\
4541.59\,\AA&1.06&3.5&3&0.5&1.14&4.5&4&0.5&1.1\\
4685.90\,\AA&1.07&2.5&2&0.5&1.2&3.5&3&0.5&1.14\\
4859.32\,\AA&1.06&3.5&3&0.5&1.14&4.5&4&0.5&1.1\\
5411.51\,\AA&1.06&3.5&3&0.5&1.14&4.5&4&0.5&1.1\\
6560.09\,\AA&1.06&3.5&3&0.5&1.14&4.5&4&0.5&1.1\\
\hline
  \end{tabular}
\end{center}
\end{table*}

\begin{table*}
\caption{Mean longitudinal field strengths (in gauss) measured from all lines
  that satisfied our 10\% below the continuum criterion. No
  entry means the line is either not present, or does not satisfy the
  criterion.}
\label{tab:results}
\begin{center}
\begin{tabular}{ccccccc}
Line & UVO\,0512--08 & CPD\,$-64^\circ 481$ & PG\,0909+276 & HD\,76431 &
 CD\,$-31^\circ 4800$ & LSE\,153 \\
 & $B_z$ (G) & $B_z$ (G) & $B_z$ (G) & $B_z$ (G) & $B_z$ (G) & $B_z$ (G) \\
\hline
H$\beta$ & $-826\pm338$ & $-1208\pm266$ & $-2596\pm475$ & $-1405\pm175$ & --- & --- \\
H$\gamma$ & $-2024\pm603$ & $201\pm406$ & $-1095\pm550$ & $-1021\pm178$ & --- & --- \\
He\,\textsc{i} 4471 & $-1915\pm533$ & --- & $-1323\pm480$ & $-1293\pm293$ & $-1356\pm324$ & --- \\
He\,\textsc{i} 4921 & $-718\pm415$ & --- & $-478\pm851$ & $-1083\pm435$ & $-1665\pm454$ & --- \\
He\,\textsc{i} 5015 & --- & --- & --- & --- & $-481\pm472$ & --- \\
He\,\textsc{i} 5875 & $-1397\pm379$ & $-718\pm745$ & $-552\pm535$ & $-650\pm170$ & $-545\pm357$ & --- \\
He\,\textsc{i} 4713 & --- & --- & --- & --- & $-1562\pm468$ & --- \\
He\,\textsc{ii} 4686 & $-1473\pm362$ & --- & $-750\pm926$ & --- & $-753\pm418$ & $-1329\pm432$ \\
He\,\textsc{ii} 5412 & --- & --- & --- & --- & $125\pm641$ & $732\pm466$ \\
He\,\textsc{ii} 4859 & --- & --- & --- & --- &  $-1016\pm941$ & $-1678\pm418$ \\
He\,\textsc{ii} 4541 & --- & --- & --- & --- & $-2692\pm951$ & $-2047\pm472$ \\
He\,\textsc{ii} 4339 & --- & --- & --- & --- & $-359\pm1376$ & $-1563\pm1103$ \\
He\,\textsc{ii} 4200 & --- & --- & --- & --- & $-1617\pm1640$ & $-1013\pm904$ \\
\hline
\textbf{Average} & $-1306\pm161$ & $-885\pm207$ & $-1448\pm222$ & $-1096\pm91$
 & $-1050\pm161$ & $-1128\pm212$ \\
\hline
\end{tabular}
\end{center}
\end{table*}

\section{Land\'e Factors}
\label{sec:lande}

Previous spectropolarimetric studies using Balmer or He\,\textsc{ii} lines
have not needed to consider the effect of the Land\'e $g$-factor on the
magnetic field strength. This is because for these lines
$g_{\mathrm{eff}}$ is approximately unity. This is not the case, however, for
many of the He \textsc{i} lines that we can measure in our
subdwarfs. We have determined the effective Land\'e factors with the classical
formula \citep[e.g.][]{LCC00}.
\begin{eqnarray}
g_{\mathrm{eff}} & = & 0.5(g_{\mathrm{low}}+g_{\mathrm{up}})+0.25(g_{\mathrm{low}}-g_{\mathrm{up}})\times
\nonumber \\
 & &
(J_{\mathrm{low}}(J_{\mathrm{low}}+1)-J_{\mathrm{up}}(J_{\mathrm{up}}+1))
\label{eq:geff}
\end{eqnarray}
where $g_{\mathrm{low}}$ and $g_{\mathrm{up}}$ are the Land\'e factors of the
lower and upper energy level. As far as the line is an electric
dipole line this formula holds independently of the coupling scheme (L-S,
j-j or intermediate coupling) for the atomic levels involved. For
calculating  $g_{\mathrm{low}}$ and $g_{\mathrm{up}}$ we used the approximation
valid for L-S coupling:
$$ g=1+\frac{J(J+1)-L(L+1)+S(S+1)}{2J(J+1)} $$
where $J$, $L$, and $S$ are the total, the orbital, and
the spin angular momentum quantum numbers of the respective energy levels. %
For hydrogen Balmer lines $g_{\mathrm{eff}}=1$ was used, although we are aware
that it is slightly different from unity. However, the deviation from unity is
less than 7\% for all lines of the Balmer series \citep{CLdI94}
which is negligible for our purposes. For the same reason, we set
$g_{\mathrm{eff}}=1$ for the He\,\textsc{ii} lines. The Land\'e factors of all
helium lines observed are shown in Table \ref{tab:lande}, along with three
strong lines that may be observable using a redder \emph{FORS1} grism. Note
that all He\,\textsc{i} transitions that have $S=0$ have $g_{\mathrm{eff}}=1$,
while $S=1$ transitions have $g_{\mathrm{eff}}>1$.

\begin{table*}
\caption{For $B_z=0$, $600$, and $1200$\,G we produced 1000 simulated
  polarisation spectra with the same noise level as the programme stars and
  used the same spectral lines as in the fits of Table \ref{tab:results}. We
  list the weighted means and standard deviations (from all spectral lines) of
  the fits at each input $B_z$ and additionally the smallest and largest
  value from all fits.}
  \label{tab:artificial}
  \begin{center}
  \begin{tabular}{lrrrrrr}
  input $B_z$ & UVO\,0512--08 & CPD\,$-64^\circ 481$ & PG\,0909+276 & HD\,76431 &
   CD\,$-31^\circ 4800$ & LSE\,153 \\
 (G) &  (G) & (G) & (G) & (G) & (G) & (G) \\
    \hline
    $0$, average & $-5\pm188$      & $3\pm300$ & $18\pm266$ & $4\pm 128$ & $0\pm 163$ & $-1\pm 239$  \\
    $0$, smallest& $-513$          & $-862$    & $-792$     & $-446$     &$-554$      & $-685$  \\
    $0$, largest & $646$           & $+831$    & $+1035$    & $+393$     &$+540$      & $+857$ \\
    $600$, average & $550\pm184$   & $582\pm294$ & $555\pm269$ & $578\pm 125$ & $539\pm 170$ & $603\pm243$  \\
    $600$, smallest& $-62$        & $-387$    & $-249$     & $+233$     &$-699$      & $-250$  \\
    $600$, largest & $+1108$       & $+1692$    & $+1603$    & $+964$     &$+1178$      & $+1346$ \\
   $1200$, average & $1106\pm186$  & $1170\pm291$ & $1119\pm266$ & $1131\pm 130$ & $1090\pm 170$ & $1194\pm243$  \\
   $1200$, smallest& $+490$        & $+164$    & $+185$     & $614$     &$+507$      & $+489$  \\
   $1200$, largest & $+1740$       & $+2196$    & $+2016$    & $+1505$     &$+1786$      & $1842$ \\
    \hline
\hline
\end{tabular}
\end{center}
\end{table*}

\section{Results}
\label{sec:results}

The results of our analysis are shown in Table \ref{tab:results} with results
for each line measured, along with the weighted average field
strength. The values shown here must be intrinsic to the star (and not
  e.g.\ interstellar or instrumental), since   no polarisation effects are
  visible in the continuum. The line errors are simply a product of the
$\chi^2$ test (see \cite{AJN04} and \cite{JWO04} for details), while the
error in the average field strength was calculated using $\sigma=\sum
1/\sigma_i^{-1/2}$, where the $\sigma_i$ are the individual line
errors. To the untrained eye, it appears that the results for
  individual lines are internally inconsistent. This will be discussed below,
  as will the reality of our errors, and the statistical significance of
field strengths we have measured.

\subsection{Statistical significance of our measurements}
\label{sec:statsig}

As was already demonstrated by \citet{AJN04} for white
dwarfs and \citet{JWO04} for central stars of planetary nebulae,
even a small polarisation signature, slightly lower than the
noise level of the spectro-polarimetric data, can be detected when
several spectral lines (and multiple observations) are used.
We again investigated this with simulated data of the
same noise level than in the subdwarf observations.

For this purpose we produced simulated polarisation spectra for
assumed magnetic fields between $0$\,G and $+1200$\,G and steps of
$200$\,G. To these polarisation
spectra we added Gaussian noise with the same standard deviation
as in the observations of the respective subdwarfs. 
We assume Gaussian noise since we expect the CCDs
  to be linear and that photon noise is the dominating limitation. The
  difference between Poisson and Gaussian noise is negligible at such high
  S/N, so we believe that our approximation is justified on theoretical
  grounds. We have also verified this by examining the frequency distribution
  of the noise.
For each of the assumed magnetic fields we made 1000 simulations and
calculated the magnetic fields of the fits from the same spectral lines we
have used for the observations of each star. In Table\,\ref{tab:artificial} we
list the results for $0$\,G, $+600$\,G, and $+1200$\,G and provided the mean
result, the standard deviation, and the smallest and largest value. This
should give us an impression of how realistic our statistical errors from the
$\chi^2$ analyses are. In this table we present only the mean field
  strength derived from our simulations, not the value for individual
  lines. It is important to note, however, that the apparently internally
  inconsistent line-by-line results are also seen in our simulations. This
  occurs because $(dI/d\lambda)/I$ in equation \ref{eq:weak} is nonlinear,
  meaning the noise deviations do not translate linearly into $V/I$
  deviations. Put simply, the weighted average field strengths are more
  well defined.

Firstly, the mean results of the respective 1000 simulations are very
close to the predescribed magnetic fields with deviations up to $100$\,G
for an assumed value of $1200$\,G. Secondly, the standard deviations are
close to the $1\sigma$ errors determined from the $\chi^2$ fits.

\begin{figure}
\vspace{11cm}
\begin{center}
    \includegraphics{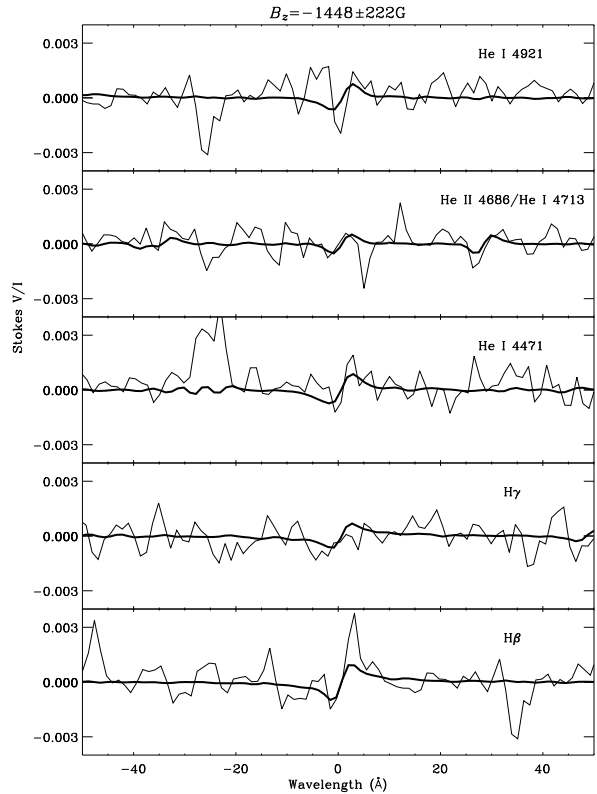}
\end{center}
\caption{Observed Stokes \emph{V/I} spectra for PG\,0909+276 for various
  hydrogen and helium lines, with value predicted by Equation \ref{eq:weak}
  overlaid (thick line). Note that the He\,\textsc{i} 4713\,\AA\ line (second
  panel from the top) has a Land\'e factor $\sim$20\% larger than the
  He\,\textsc{ii} 4686\,\AA\ line.}
\label{fig:pg0909}
\end{figure}

\begin{figure}
\vspace{11cm}
\begin{center}
    \includegraphics{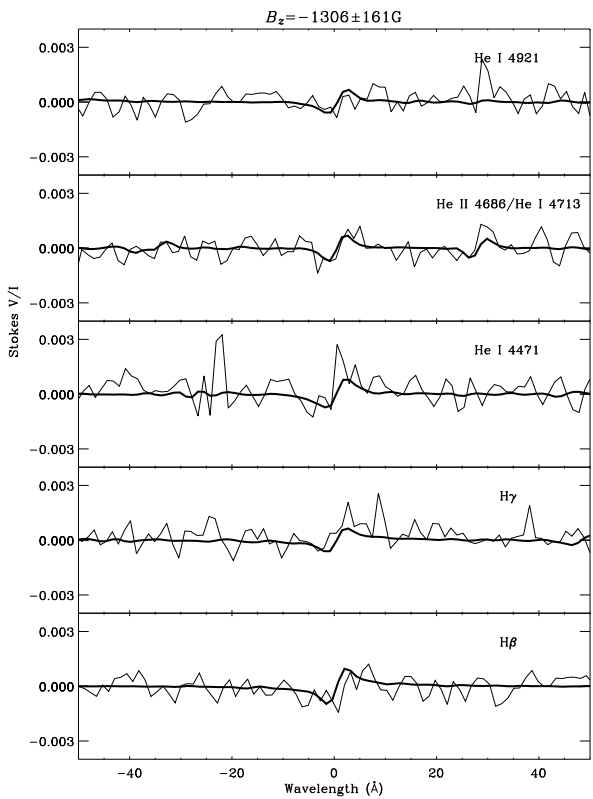}
\end{center}
\caption{Similar to Figure \ref{fig:pg0909}, except for UVO\,0512--08.}
\label{fig:uvo0512}
\end{figure}

\begin{figure}
\vspace{6.5cm}
\begin{center}
    \includegraphics{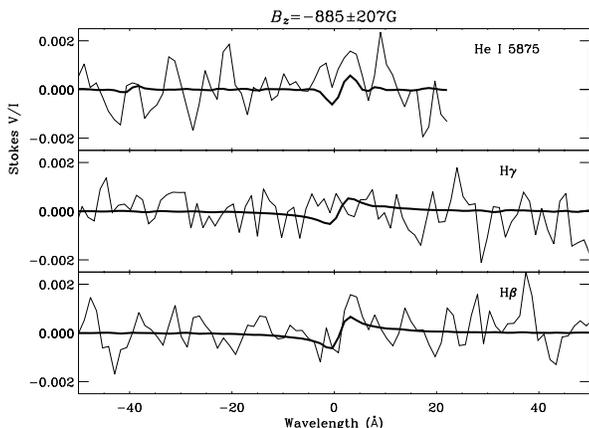}
\end{center}
\caption{Similar to Figure \ref{fig:pg0909}, except for CPD\,$-64^\circ
  481$. The top panel shows the He\,\textsc{i} 5875\,\AA\ line at the edge of
  the spectrum.}
\label{fig:cpdm64}
\end{figure}

\begin{figure}
\vspace{11cm}
\begin{center}
    \includegraphics{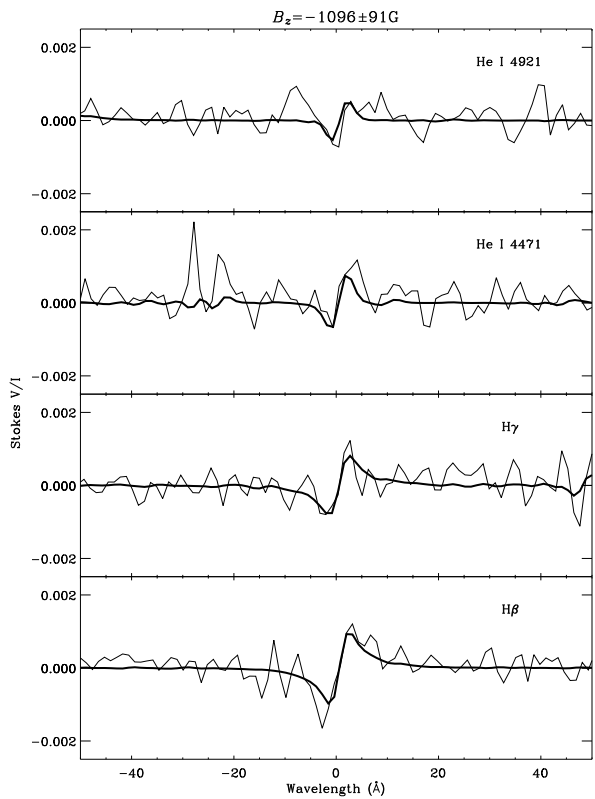}
\end{center}
\caption{Similar to Figure \ref{fig:pg0909}, except for HD\,76431. In this
  star the polarisation signature is evident in every line.}
\label{fig:hd76431}
\end{figure}

\begin{figure}
\vspace{11cm}
\begin{center}
    \includegraphics{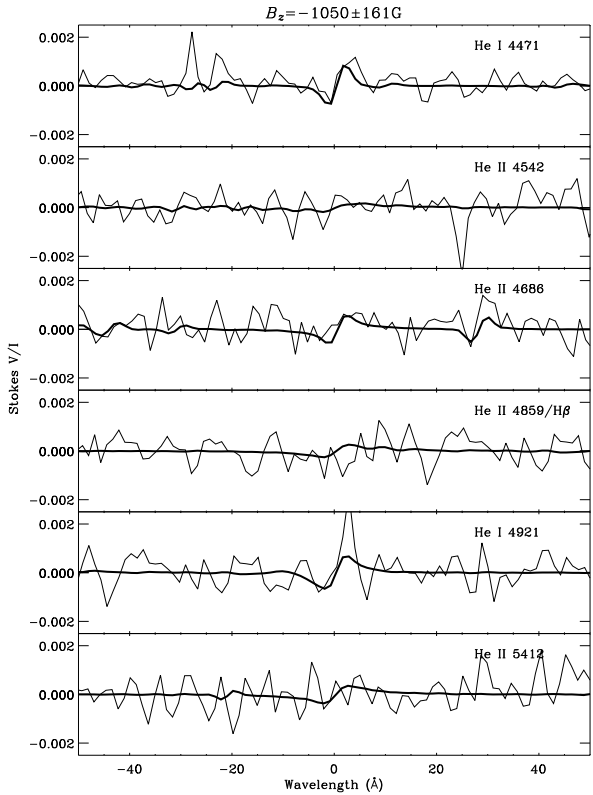}
\end{center}
\caption{Similar to Figure \ref{fig:pg0909}, except for CD\,$-31^\circ 4800$.}
\label{fig:cdm31d4800}
\end{figure}

\begin{figure}
\vspace{11cm}
\begin{center}
    \includegraphics{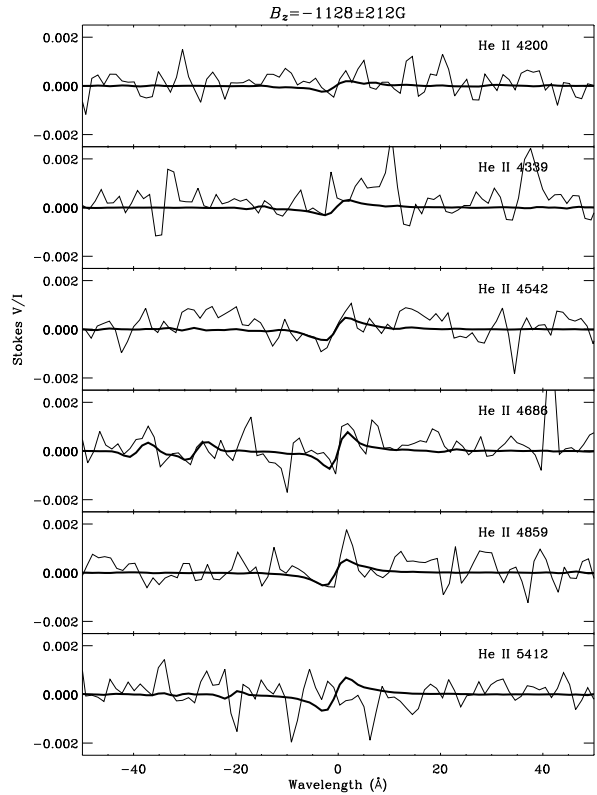}
\end{center}
\caption{Similar to Figure \ref{fig:pg0909}, except for LSE\,153.}
\label{fig:lse153}
\end{figure}

Moreover, the smallest and largest values from all 1000 fits for
an assumed field of $0$\,G  provides
a very conservative estimate of what magnetic field can be mimicked by pure
noise: in all cases the fitted field strengths are larger in their
absolute value than these extreme numbers; only in the case of
CPD\,$-64^\circ 481$ the derived magnetic field strength of
$B_z=-885$\,G comes close to the simulated extreme deviation from zero
of $-862$\,G. Therefore, we conclude that all of our results
are incompatible with $B_z=0$, with a somewhat smaller significance for
CPD\,$-64^\circ 481$. Two of the objects (UVO\,0512--08 and  HD\,76431) are
even incompatible with $B_z=600$\,G. This indicates that our $1\sigma$ errors
are very realistic and that all stars very probably have a magnetic
field. Note also that our error ranges are much smaller than for the stars
studied by both \citet{AJN04} and \citet{JWO04}, in part because we often have
more spectral lines available to measure and because some of the objects
are much brighter. The individual objects are discussed in more detail
below, and their Stokes \emph{V/I} spectra are shown in Figures
\ref{fig:pg0909}--\ref{fig:lse153}.

\subsection{Super-metal-rich sdBs}
\label{sec:metalrich}

These stars are the super-metal-rich sdOB objects discovered by
\citet{EHN01}, so we are particularly interested in their magnetic field
strengths. As can be seen from Table \ref{tab:results}, we measure
$B_z=-1306\pm161\,\mathrm{G}$ for UVO\,0512--08 and
$B_z=-1448\pm222\,\mathrm{G}$ for PG\,0909+276. Both of these results are
statistically significant (see the previous section), and are also the two
highest field strengths we have measured. Whether or not this is physically
significant remains to be seen, since the two ``normal'' sdBs we have observed
may have evolved differently (as a binary system), or have evolved well away
from the EHB. The super-metal-rich stars are both apparently single, although
with its low surface gravity and high temperature, UVO\,0512--08 appears to
have also evolved away from the EHB.

\subsection{CPD\,$\mathit{-64^\circ 481}$ -- a typical sdB star}
\label{sec:normal}

CPD\,$-64^\circ 481$ is an sdB in a short-period (0.27\,d) binary with an
unseen companion of low mass, possibly a brown dwarf (Edelmann et al., in
preparation). This star has the lowest measured field and the measurement with
the lowest significance, with $B_z=-885\pm207\,\mathrm{G}$, however,
the probability for an almost kilogauss field is still very
high. CPD\,$-64^\circ 481$ had the smallest number of spectral lines available
to measure polarisation. We are uncertain how binary evolution -- most likely
the system has passed through a common envelope phase -- might affect the
magnetic field of the star's progenitor.

\subsection{HD\,76431 -- an evolved sdB star}
\label{sec:evolved}

HD\,76431 was originally classified as a main sequence star at high
  galactic latitude. Its subluminous nature was discovered by a detailed
  quantitative spectral analysis by \citet{RHE01}. The helium deficiency
  and a peculiar metal abundance is similar to that of the sdB stars, although
  the deviation from solar values are smaller than typical for these
  stars. The gravity of HD\,76431 is slightly higher than that of a main
  sequence star, but considerably lower than for EHB stars. Comparison to
  EHB evolutionary tracks suggests that it has already left the EHB and is
  evolving towards the white dwarf cooling sequence as has been suggested for
  some sdO stars. Therefore HD\,76431 is not a typical sdB, but might be
  considered as a link between the sdB and sdO stars. The magnetic field
  measured ($B_Z=-1096\pm91$\,G) has very high significance as found in
  Section \ref{sec:statsig}. This star has the clearest ``by eye''
  polarisation signature.

\subsection{Helium-rich sdOs}
\label{sec:sdOs}

Both of the sdO stars we observed are helium-rich. For CD\,$-31^\circ 4800$
\citet{BH95} found $T_{\mathrm{eff}}=$\,44\,000\,K and $\log g=$\,5.9 and
measured abundances; the star contains $\sim30$ time more helium then
hydrogen, and appears to be a ``garden variety'' sdO \citep[see,
  e.g. ][]{SHL05}. We measure $B_z=-1050\pm161\,\mathrm{G}$ based on the 11
lines shown in Table \ref{tab:results}, and find from our simulations that
this value is statistically significant.

LSE\,153 is an extremely hydrogen-deficient sdO that has been suggested as a
possible successor of a R CrB star, with a mass of $\sim$0.7\,$M_\odot$
\citep{HBH89}. The star's very low hydrogen abundance is accompanied by both
nitrogen and carbon enhancements. The field strength we measure
($B_z=-1128\pm212$\,G) was also found to be statistically significant. This
star is at a similar evolutionary stage as the central stars of planetary
nebulae studied by \citet{JWO04}, and has a field strength of approximately
the same order of magnitude.

In both of the helium-rich sdOs we have observed the magnetic field determined
using the He\,\textsc{ii} 5412 line does not match the other He\,\textsc{ii}
lines. We cannot see any obvious reason for this strange behaviour.

\section{Discussion}
\label{sec:discuss}

The goal of our project was to determine whether magnetic field strength is
directly connected to abundance anomalies in these stars. The magnetic field
could have an influence on the chemical composition by suppressing convection
or by influencing mass-loss or diffusion. While convection is not expected to
be important in helium-poor sdB stars, He\,\textsc{ii} ionisation can initiate
convection in helium-rich sdO stars at $T_{\mathrm{eff}}$ between 40\,000 and
50\,000\,K \citep{GKH85}. This may be relevant for CD\,$-31^\circ 4800$, but
not for the other programme stars. However, no clear differences exist between
the field strengths of the "normal" and super metal-rich stars in our sample.

The origin of the fields detected here are unknown, and  are found in
objects from various evolutionary channels. It is possible that they are
dynamo induced or a remnant from their progenitor. We consider a
dynamo-induced magnetic field may be possible if rapid internal rotation
occurs as suggested recently by \citet{KH05}, who found evidence that
several of the sdB pulsators may have rapidly rotating cores. This is
contrasted with the stars' slow surface rotation \citep[typically $v\sin i\le
  5$\,km\,s$^{-1}$, e.g. ][]{HRW00}. While the sdBs have probably followed a
similar evolutionary path, the same cannot be said for at least one of the sdO
stars. LSE\,153 can be associated with the "born-again" post-AGB scenario
\citep{HBH89}, while the helium-rich sdO, CD\,$-31^\circ 4800$, is similar to
most other objects of its class.

It is interesting to estimate the strength of the magnetic fields when these
stars have evolved to white dwarfs. If we assume complete conservation of
magnetic flux the magnetic fields are amplified inversely proportional to the
square of the stellar radius. The radii of the sdB stars are
0.15--0.25\,$R_\odot$ (taking the canonical mass of an EHB star to be
0.48\,$M_\odot$). Since they will evolve directly into white dwarfs, their
radii will shrink by a factor of $\sim$20, leading to field strengths of up to
500\,kG, values which are apparently rarely seen. The search for rotation in
white dwarfs \citep{HNR97,KDW98,KNH05} from high resolution H$\alpha$
spectroscopy also resulted in constraints on field strengths. Amongst the
$\sim$50 DA white dwarfs studied, only four turned out to be magnetic with $B$
up to 180\,kG, while upper limits of 10-20\,kG could be derived for the rest.
\cite{AJN04} have detected magnetic fields between 2000 and 4000\,G in four
white dwarfs, while no magnetic fields on this level of accuracy was found in
the other eight programme stars.
However, since subdwarfs contribute to only a small percentage (1-2\%) of
white dwarfs progenitors, and the current sample of sub-MG white dwarfs is
still small, we cannot completely rule out the existence of low-mass ($\sim0.5
M_\odot$) magnetic white dwarfs, although the population must not be
large. It is much more difficult to work backwards to values one might
  expect for the main sequence progenitors of our targets, since sdB formation
  is not well defined. However, if we take a star with the same parameters as
  the Sun as an example, we find field strengths in the 20-40\,G range, which
  seem reasonable (although currently undetectable). This also appears to rule
  out any evolutionary connection with the Ap stars.

\citet{JWO04} have found kilogauss magnetic fields in all of the central stars
of planetary nebula that they have observed by means of
spectro-polarimetry. By magnetic field amplification these stars will have
magnetic fields strengths between 0.35 and 2\,MG if the assumption of complete
conservation of magnetic flux is true. The same holds for the helium-rich sdO
star LSE\,153 analysed in this paper. Since about two thirds of the white
dwarfs seems to have magnetic field even lower than 1\,kG \citep{AJN04} one
must conclude that magnetic fields can be destroyed during the final stages of
stellar evolution, although the detailed mechanisms are unknown.

There are several questions that have arisen from our detection of magnetic
fields in hot subdwarfs. Firstly, are the field strengths dependent on the
binary status of the star? The one star we have observed that is known to be
in a binary system has the lowest and a somewhat less statistically significant
measurement. Since it is only one object we cannot make any conclusions at
this stage, but this may warrant further investigation. Secondly, what effect,
if any, might kilogauss magnetic fields have on pulsations in sdB stars?
Rotation is often considered in connection to sdB pulsators, however magnetic
fields have not as yet. Asteroseismology has been used to infer a magnetic
field in at least one white dwarf \citep[the pulsating DB GD\,358, see
][]{WNC94}. This field may be dynamo induced
\citep{MTvH94}. Spectropolarimetric observations of sdB pulsators, along with
a theoretical investigation, would help to clarify this situation.

\section{Conclusions}

In this study, we have used polarisation measurements of sdB and sdO stars to
try to determine whether magnetic fields in two super-metal-rich stars can
explain their extreme abundance peculiarities. Field strengths of up to
$\sim$1.5\,kG range have been measured at varying levels of significance in
each of our six targets, however no clear difference was found between
apparently normal subdwarfs and the metal-rich objects. The origin of the
magnetic fields is unknown. We also considered the implications of our
measurements for magnetic flux conservation in late stages of evolution.
Unless there is a population of low-mass white dwarfs with field strengths up
to $\sim$\,500\,kG, it is unlikely that magnetic flux is conserved in late
stages of evolution. This idea is consistent with findings for central stars
of planetary nebulae.

\acknowledgements
We would like to thank E.\ degl'Innocenti for his advice on Land\'e
factors, and Svetlana Hubrig for help with the observational setup. SJOT is
supported by the Deutsches Zentrum f\"ur Luft- und Raumfahrt (DLR) through
grant no.\ 50 OR 0202.

\bibliographystyle{aa}

\end{document}